\documentstyle[sprocl]{article}

\def\Journal#1#2#3#4{{#1} {\bf #2}, #3 (#4)}

\def\PRD{{\em Phys. Rev.} D}

\def\CQG{\em Class. Quantum Grav.}
\def\MNRAS{\em Mon. Not. R. Astr. Soc.}

\def\be{\begin{equation}}
\def\ee{\end{equation}}
\def\bea{\begin{eqnarray}}
\def\eea{\end{eqnarray}}

\begin{document}

\title{COVARIANT COSMOLOGICAL PERTURBATION DYNAMICS IN THE INFLATIONARY UNIVERSE}

\author{W. ZIMDAHL}

\address{Fakult\"at f\"ur Physik, Universit\"at Konstanz, PF 5560 M678, \\
D-78457 Konstanz, Germany\\
E-mail: winfried.zimdahl@uni-konstanz.de} 


\maketitle
\abstracts{The linear cosmological perturbation theory of almost homogeneous and isotropic perfect fluid and scalar field universes is reconsidered and formally simplified. 
Using the existence of a covariant conserved quantity on large perturbation scales, a closed integral expression for comoving energy density perturbations is obtained for arbitrary equations of state. 
On this basis we establish a simple relation between fluid 
energy 
density perturbations at `reentry' into the horizon and the corresponding 
scalar field quantities at the first Hubble scale crossing 
during an early de Sitter phase of a standard inflationary scenario.}

\section{Introduction}
According to the standard inflationary picture the presently observed large-scale structures in the universe may be traced back to quantum fluctuations during an early de Sitter phase. 
These originally
small-scale perturbations are stretched out tremendously in the inflationary
period, thereby crossing the Hubble length and becoming large-scale perturbations. Lateron, after the
inflation has finished, i.e., when the universe is adequately described by
the standard Friedmann-Lema\^{\i}tre-Robertson-Walker (FLRW) model, these
perturbations cross the Hubble length for the second time, now inwards, and again
become small-scale perturbations. 

The inhomogeneities now carry an imprint of the fluctuations then. 
In order to establish a connection between perturbation amplitudes during an early inflationary phase  and cosmological structures observed today it is of interest to follow the perturbation dynamics during the time interval between both Hubble scale crossings. 
Assuming the perturbations to behave classically throughout, the present contribution addresses this problem within a covariant approach. 
A covariant approach which uses exactly defined tensorial quantities instead of conventional, generally gauge-dependent perturbation variables, is conceptionally superior to noncovariant approaches since it avoids the explicit introduction of a fictitious background universe. 
It thus circumvents the gauge problem which just originates from the nonuniqueness of the conventional splitting of the spacetime into a homogeneous and isotropic zeroth order and first-order perturbations about this background. 

We devote special attention to a unified description of scalar fields and perfect fluids and to the characterization of a covariant conserved quantitiy on large perturbation scales. 
The existence of such a quantitiy allows us to encode the entire linear perturbation dynamics in a closed time integral and to find a first-order relation between comoving fluid energy density perturbations at ``reentry'' into the horizon and corresponding scalar field energy density perturbations at the first Hubble scale crossing during an early de Sitter phase. 

In section 2 we reconsider the linear cosmological perturbation theory of almost homogeneous and isotropic perfect fluid or scalar field universes. 
A formal simplification of the dynamical description is obtained 
in terms of a new covariant perturbation variable, representing fractional energy density perturbations on hypersurfaces of constant Hubble parameter, in section 3.  
Use of this variable allows us to identify a conserved quantity on large perturbation scales 
in the spatially flat case and to solve the corresponding perturbation dynamics in section 4.  
On this basis we find, in section 5, the comoving fluid energy density perturbations at reentry into the horizon in terms of the corresponding scalar field quantity at the first Hubble scale crossing during an early slow-roll phase. \\
This contribution summarizes recent work on covariant cosmological perturbation theory published elsewhere \cite{Z1,Z2,ZP}. 

\section{Basic covariant perturbation dynamics}
We consider cosmic media characterized
by an energy-momentum tensor 
\begin{equation}
T _{mn} = \rho u _{m}u _{n} + p h _{mn}\ ,
\mbox{\ \ \ \ }
\left(m,n... = 0,1,2,3 \right)\ ,
\label{1}
\end{equation}
where $\rho $ is the energy density, $p$ is the pressure, $u_{m}$  
is the four-velocity ($u ^{m}u _{m} = -1$), and
$h_{mn} = g_{mn} + u_{m}u_{n}$. 
We recall that the structure (\ref{1}) is also valid for a  
minimally coupled scalar field with the identifications
\begin{equation}
\rho  = \frac{1}{2}\dot{\phi }^{2} + V \left(\phi  \right)\ ,
\mbox{\ \ \ \ }
p  = \frac{1}{2}\dot{\phi }^{2} - V \left(\phi  \right)\ ,
\mbox{\ \ \ \ }
u _{i} =  - \frac{\phi _{,i}}
{\sqrt{-g ^{ab}\phi _{,a}\phi _{,b}}}\ ,
\label{2}
\end{equation}
where $\dot{\phi } \equiv  \phi _{,a}u ^{a} = \sqrt{-g ^{ab}\phi  
_{,a}\phi _{,b}}$ and $V \left(\phi  \right)$ is the scalar field  
potential. 

The key quantity of the covariant aproach is the covariantly defined spatial derivative of the energy density, $h ^{c}_{a}\rho _{,c}$. 
This variable which vanishes in a homogeneous universe  
may be used to describe inhomogeneities in the energy density without explicitly introducing a fictitious background universe. 
$h ^{c}_{a}\rho _{,c}$ is an exactly defined tensorial quantity and no perturbation variable in the usual sense. 
In order to clarify its relation to conventional perturbation quantities we may, however, decompose it into a zeroth (superscript (0)) 
and a first order (superscript (1))  
as though it were a usual perturbation variable. 
Assuming a homogeneous and isotropic comoving zeroth order we find in such a case  $\left(\mu ... = 1,2,3 \right)$ 
\begin{equation}
\left(h ^{c}_{a}\rho _{,c}\right)^{^{\left(0 \right)}} = 0 
 \ ,\ \ \ \ \ 
\left(h
^{c}_{0}\rho _{,c}\right)^{^{\left(1 \right)}} 
= 0
\ \ \ \ \ ,
\left(h
^{c}_{\mu }\rho _{,c}\right)^{^{\left(1 \right)}} 
= \rho ^{^{\left(1 \right)}}_{,\mu } + \dot{\rho }
^{^{\left(0 \right)}}u ^{^{\left(1 \right)}} _{\mu } \ .  
\label{3}
\end{equation}
Under
infinitesimal coordinate transformations 
$x ^{n \prime} = x ^{n} - \zeta ^{n}\left(x\right)$ 
the perturbation quantities in (\ref{3}) transform as 
\begin{equation}
\rho ^{^{\left(1 \right)\prime }} = \rho ^{^{\left(1 \right)}} + \dot{\rho }^{^{\left(0 \right)}}\zeta
^{0}\ ,\ \ \ u_{\mu }^{^{\left(1 \right)\prime }} 
= u ^{^{\left(1 \right)}}_{\mu } - \zeta ^{0}_{,\mu }\ . 
\label{4}
\end{equation}
It is obvious, that the combination on the right-hand side of the last equation (\ref{3}) is gauge-invariant. 
The point of view of conventional perturbation theory is to put together 
two gauge-dependent quantities, in the present case 
$\rho ^{^{\left(1 \right)}}$ and $u ^{^{\left(1 \right)}}_{\mu }$, such that the combination is invariant. 
Working with a covariant quantity, a corresponding problem does not appear. 
If treated as a usual first-order perturbation variable, the gauge-invariance of the resulting expression is guaranteed. 
This illustrates the conceptional advantage of the covariant approach. 
Incidentally, $\left(h^{c}_{\mu }\rho _{,c} \right)^{^{\left(1 \right)}}$ is  the gradient of the quantity $\epsilon _{m}$ 
used by Bardeen \cite{Bardeen}. 

Suitable fractional variables to characterize spatial inhomogeneities are 
\cite{Z1,Z2,ZP,EB,Jack}
\begin{equation}
D _{a} \equiv \frac{S h ^{c}_{a}\rho _{,c}}{\rho + p} \ ,
\mbox{\ \ \ \ }
P _{a} \equiv \frac{S h ^{c}_{a}p _{,c}}{\rho + p} \ ,
\mbox{\ \ \ \ }
t _{a} \equiv S h ^{c}_{a} \Theta _{,c} \ ,
\label{5}
\end{equation}
where $S$ is a length scale generally defined by $\Theta \equiv  3  
\dot{S}/S$.
The quantities $D _{a}$ and $P _{a}$ represent fractional, comoving  
(with the fluid four-velocity) energy density and pressure  
perturbations, respectively.
Inhomogeneities in the expansion are described by the quantity $t _{a}$. 

The dynamics of the inhomogeneities may be obtained within the so-called 
``fluid-flow'' approach by using the energy-momentum conservation  
$T ^{ik}_{\ ; k} = 0$, implying 
\begin{equation}
\dot{\rho } = - \Theta\left(\rho + p\right) \ ,
\mbox{\ \ \ \ }
\left(\rho  + p\right)\dot{u}_{}^{m} = 
- p_{,k}h^{mk}
\label{6}
\end{equation}
and the Raychaudhuri equation
\begin{equation}
\dot{\Theta} + \frac{1}{3}\Theta^{2} 
+ 2\left(\sigma^{2} - \omega^{2}\right) - \dot{u}^{a}_{;a} 
- \Lambda 
+ \frac{\kappa}{2}\left(\rho  + 3p\right) 
= 0 \ . 
\label{7}
\end{equation}
Differentiating the energy balance in (\ref{6}), projecting orthogonal to $u _{a}$, multiplying by $S$ and restricting ourselves to first-order deviations from homogeneity and isotropy, we obtain 
\begin{equation}
\dot{D}_{\mu }
+ \frac{\dot{p}}
{\rho  + p} D _{\mu }
+ t _{\mu } = 0 \ .
\label{8}
\end{equation}
In a similar way a corresponding equation for $t _{\mu }$ follows from (\ref{7}). 
Eliminating $t _{\mu }$ from this system we find 
\begin{eqnarray}
&&\ddot{D}_{\mu } + \left(\frac{2}{3} - c _{s}^{2}\right)
\Theta \dot{D}_{\mu }
- \left[\left(c_{s}^{2} \right)^{\displaystyle \cdot}\Theta 
\right.\nonumber\\
&&+ \left. \left(\frac{\kappa }{2}\left(\rho - 3 p\right) + 2 \Lambda\right)
c _{s}^{2} + \frac{\kappa }{2} \left(\rho + p\right) \right] D _{\mu } 
=  \frac{\nabla ^{2}}{a ^{2}} P _{\mu }\ ,
\label{9}
\end{eqnarray}
where $c _{s}^{2} = \dot{p}/\dot{\rho }$ and $S$ has reduced to the scale factor $a$ of the Robertson-Walker metric. 
For a fluid (superscript f) one has
$P _{a}^{^{\left(f \right)}} = c_{s}^{2} D _{a}^{^{\left(f \right)}}$
and Eq. (\ref{9}) corresponds to Jackson's \cite{Jack} equation (57). 
For a scalar field
(superscript s),  because of $h ^{c}_{a}\phi _{,c} = 0$, the  
potential term neither contributes to $D _{a}^{^{\left(s \right)}}$  
nor to
$P _{a}^{^{\left(s \right)}}$ and, consequently,
$P _{a}^{^{\left(s \right)}} =  D _{a}^{^{\left(s \right)}}$
is valid \cite{Z2} which, if used in Eq. (\ref{9}), results in a  
closed equation for
$D _{\mu }$ as well.

\section{A new perturbation variable}
Let us now consider again relation (\ref{3}) which was used to establish the connection between the covariant quantity $h ^{c}_{m}\rho _{,c}$ and conventional perturbation variables. 
By  suitable infinitesimal coordinate transformations each term on the right-hand side of the last relation (\ref{3}) may be separately transformed away but not the combination. 
In the special case $u ^{^{\left(1 \right)}}_{\mu }=0$ (comoving hypersurfaces) the quantity 
$h^{c}_{\mu }\rho _{,c}$ in first order corresponds to a pure energy density perturbation. 
Consequently, it is interpreted as first-order energy density perturbation on comoving hypersurfaces. 
As to the physical interpretation of the gauge-invariant quantity 
$h ^{c}_{\mu }\rho _{,c}$ the comoving gauge seems preferred. 
Of course, there is no reason to wonder about this since $\rho $ is defined as the energy density of a comoving observer. 
However, it is well-known
that there exist obviously reasonable gauge-invariant quantities 
with a  physical meaning in gauges different from the
comoving gauge. 
One may therefore ask whether there are covariant and
gauge-invariant quantities which, from the point of view of their physical
interpretation, prefer other gauges in a similar sense in which the quantity 
$h ^{c}_{\mu }\rho _{,c}$ 
prefers the comoving gauge. 

Let us consider the ratio of the spatial variation of a scalar quantity,
say, the energy density, i.e. $h ^{c}_{a}\rho _{,c}$, to its variation in
time, $\dot{\rho } \equiv u ^{c}\rho _{,c}$, and the corresponding ratio for 
$\Theta $. 
In
linear order we find  
\begin{equation}
\left(\frac{h ^{c}_{\mu }\rho _{,c}}{ u ^{c}\rho _{,c}}\right)
^{^{\left(1 \right)}} = 
\frac{\rho ^{^{\left(1 \right)}}_{,\mu }}{\dot{\rho }} 
+ u ^{^{\left(1 \right)}}_{\mu }\ ,  
\mbox{\ \ \ \ }
\left(\frac{h ^{c}_{\mu }\Theta _{,c}}{ u ^{c}\Theta _{,c}}\right)
^{^{\left(1 \right)}}
= \frac{\Theta ^{^{\left(1 \right)}}_{,\mu }}{\dot{\Theta }} 
+ u ^{^{\left(1 \right)}}_{\mu }\ .  
\label{10}
\end{equation}
Obviously, these ratios are  reasonable quantites to characterize small
deviations from homogeneity. 
The 4-velocity perturbation 
$u ^{^{\left(1 \right)}}_{\mu}$ enters both expressions in (\ref{10}) in exactly the
same manner, namely simply additively. 
This suggests combining the ratios in (\ref{10}) such that the perturbations of 
the four-velocity drop out, i.e., to consider the difference 
\begin{equation}
\frac{h ^{c}_{m }\rho _{,c}}{ u ^{c}\rho _{,c}} - \frac{h ^{c}_{m }\Theta
_{,c}}{ u ^{c}\Theta _{,c}} \ .  
\label{11}
\end{equation}
This combination is an exact covariant quantity which vanishes in a homogeneous universe since each term vanishes separately. 
Consequently, the combination (\ref{11}) may be used to characterize inhomogeneities in a covariant way. 
In linear order we obtain 
\begin{equation}
\left[\frac{h ^{c}_{m }\rho _{,c}}{ u ^{c}\rho _{,c}} - \frac{h ^{c}_{m }\Theta
_{,c}}{ u ^{c}\Theta _{,c}} \right]^{^{\left(1 \right)}} 
= \frac{\rho ^{^{\left(1 \right)}}_{,\mu }}{\dot{\rho }} 
- \frac{\Theta ^{^{\left(1 \right)}}_{,\mu }}{\dot{\Theta }} \
.
\label{12}
\end{equation}
Again, each term on the right-hand side is not invariant 
separately under infinitesimal  coordinate transformations while the combination is. 
For the special choice $\Theta ^{^{\left(1 \right)}}_{,\mu }=0$ e.g., the quantity (\ref{12}) describes pure energy density perturbations. 
It follows that the combination (\ref{11}) may be interpreted as describing energy density perturbations on hypersurfaces of constant expansion in a similar sense 
in which $h ^{c}_{m }\rho _{,c}/ \dot{\rho }$ describes energy density perturbations on comoving hypersurfaces.\cite{Z1} 
The corresponding fractional quantity is defined as 
\begin{equation}
D _{a}^{\left(ce \right)} \equiv
D _{a} - \frac{\dot{\rho }}{\rho + p}
\frac{t _{a}}{\dot{\Theta }}\ .
\label{13}
\end{equation}
The superscript ($ce$) stands for ``constant expansion''. 
The essential advantage of using this quantity as basic perturbation variable is that it allows us to rewrite the entire linear perturbation dynamics (\ref{9}) in the following compact form: 
\begin{equation}
\left[a ^{2}\dot{\Theta }
D _{\mu }^{\left(ce \right)}\right]^{\displaystyle \cdot}
=   - a ^{2}\Theta \frac{\nabla ^{2}}{a ^{2}}P _{\mu } \ .
\label{14}
\end{equation}

\section{Solving the large-scale perturbation dynamics}
Except for the spatial pressure gradient on the right-hand side of
Eq. (\ref{14}) all terms of the dynamical perturbation equation have   
been included into a first time derivative. 
Obviously, the quantity 
$a ^{2}\dot{\Theta }D _{\mu }^{\left(ce \right)}$ in the bracket on the left-hand side of Eq. (\ref{14}) is approximately conserved if the spatial gradient term on the right-hand side of that equation may be considered small. 
This is the case on large perturbation scales. 
Generally, the importance of conserved large-scale quantities is well recognized in the literature (see references in \cite{Z1,Z2,ZP}).  
Using the eigenvalue structure $\nabla ^{2} \rightarrow - \nu ^{2}$ of the three-dimensional Laplacian 
in a spatially flat universe, where the continuous parameter $\nu$ is related to the 
physical wavelength $\lambda $ by $\lambda = 2 \pi a/ \nu$, we find 
\begin{equation}
a ^{2}\dot{\Theta }D _{\left(\nu \right) }^{\left(ce \right)}
\equiv   - E _{\left(\nu \right)}  =  {\rm const} \ ,
\mbox{\ \ \ \ }
\left(\nu \ll 1 \right)\ . 
\label{15}
\end{equation}
This property holds both for perfect fluids and scalar fields. 

Taking into account the relation 
between $D _{\mu }^{\left(ce \right)}$ and $D _{\mu }$ which  
follows from the definition (\ref{13}) of $D _{\mu }^{\left(ce  
\right)}$ and Eq. (\ref{8}),
the equation to solve for $D _{\left(\nu \right)}$ is a first-order differential equation instead of the second-order equation (\ref{9}), 
\begin{equation}
\dot{D}_{\left(\nu \right) } - \left(\frac{\dot{\Theta }}{\Theta }
+ c _{s}^{2} \Theta  \right)D _{\left(\nu \right)}
= \frac{E _{\left(\nu \right) }}{a ^{2}\Theta }\ ,
\mbox{\ \ \ \ \ }
\left(\nu \ll 1 \right)\ .
\label{16}
\end{equation}
With $\Theta = 3 \dot{a}/a$ and $c _{s}^{2} \approx {\rm const}$ the solution of 
(\ref{16}) becomes\cite{ZP}
\begin{equation}
D _{\left(\nu \right)} = \Theta a ^{3 c _{s}^{2}}
\left[\int^{t} dt \left(
\frac{E _{\left(\nu \right)}}{\Theta ^{2}a ^{2 + 3 c _{s}^{2}}} \right)
+ C _{\left(\nu \right)}\right]\ , 
\mbox{\ \ }
\left(\nu \ll 1 \right) \ ,
\label{17}
\end{equation}
where $C _{\left(\nu \right)}$ is an integration constant.
Formula (\ref{17}) comprises the entire large-scale linear  
perturbation dynamics for scalar fields and perfect fluids of arbitrary equations of state.

\section{Transfer function in the inflationary universe}
Let us first apply the general formula (\ref{17}) to the 
``slow-roll'' phase of a scalar field dominated universe. 
Under the corresponding condition 
$\ddot{\phi } \ll \Theta \dot{\phi }$ we have 
$\Theta = 3H = {\rm const}$ as well as $c _{s}^{2} = - 1$ and integration of (\ref{17}) yields
\begin{equation}
D ^{^{\left(s \right)}}_{\left(\nu \right)} 
= \frac{E _{\left(\nu \right)}}{3 H ^{2}a ^{2}} 
+ 3 \frac{H C _{\left(\nu \right)}}{a ^{3}}\ , 
\mbox{\ \ }
\left(\nu \ll 1 \right)\ .
\label{18}
\end{equation}
This solution describes the exponential damping of the 
comoving, fractional energy density gradient on large scales 
during the de Sitter phase. 
It demonstrates the stability of the latter with respect to linear 
perturbations and seems to 
represent a gauge-invariant formulation of the cosmic 
``no-hair'' theorem.  
While the perturbation wavelengths are stretched out tremendously, the 
amplitude of the comoving, fractional energy density gradient becomes 
exponentially small. 
As will be shown below, it is {\it not} justified, however, to neglect 
these perturbations since they will resurrect during the 
subsequent FLRW phase. 
While the ``hair'' gets extremely shortend during the 
de Sitter period, it is not completely extinguished but 
slowly grows again afterwards. 

In order to determine the transfer functions which relate fractional energy density perturbations at Hubble horizon crossing during an early de Sitter phase to corresponding perturbations at reentry into the particle horizon during a subsequent FLRW period, 
we first 
find the general solution of the integral formula (\ref{17}) in case the cosmic substratum is a fluid with 
$c _{s}^{2} \rightarrow \gamma - 1 $,  
where we assume 
$\gamma  
\equiv  1 + p /\rho $ constant. 
Under this condition we have 
$a \propto t ^{2/ \left(3 \gamma  \right)}$ and 
$H = 2/ \left( 3 \gamma t \right)$.  
The general formula (\ref{17}) then integrates to 
\begin{equation}
D ^{^{\left(FLRW \right)}}_{\left(\nu \right)} = \frac{E _{\left(\nu \right)}}{3}
\frac{2}{2 + 3\gamma}
\frac{1}{H ^{2}a ^{2}} 
+ 3 C _{\left(\nu \right)} H a ^{3 c _{s}^{2}}\ , 
\mbox{\ \ }
\left(\nu \ll 1 \right)\ .
\label{19}
\end{equation}
Introducing the abbreviations 
\begin{equation}
D _{\left(\nu \right)}^{^{\left(s,d \right)}} = 
\frac{E _{\left(\nu \right)}}{3}\frac{1}{H ^{2}a ^{2}}\ , 
\mbox{\ \ \ }
D _{\left(\nu \right)}^{^{\left(FLRW,d \right)}} = 
\frac{E _{\left(\nu \right)}}{3}\frac{2}
{3\gamma + 2}
\frac{1}{H ^{2}a ^{2}} 
\label{20}
\end{equation}
for the dominating (superscript $d$) modes in the slow-roll and the FLRW phases, respectively, the conservation property of 
$E _{\left(\nu \right)}$ 
allows us to obtain a relation between 
these quantities at different times:
\begin{equation}
D _{\left(\nu \right)}^{^{\left(FLRW,d \right)}}\left(t _{e} \right) = 
\frac{2}{3 \gamma  + 2}
\frac{H ^{2}\left(t _{i} \right)a ^{2}\left(t _{i} \right)}
{H ^{2}\left(t _{e} \right)a ^{2}\left(t _{e} \right)}
D _{\left(\nu \right)}^{^{\left(s,d \right)}}\left(t _{i} \right) 
\mbox{\ \ }
\left(\nu \ll 1 \right)\ .
\label{21}
\end{equation}
Here, $t _{i}$ is the time of the initial horizon crossing during the de Sitter phase while 
$t _{e}$ denotes the time of reentry during the subsequent FLRW phase.
The horizon crossing conditions for a constant, comoving wavelength 
$\lambda /a$ are  
\begin{equation}
\frac{\lambda }{a} = 
\left[H \left(t _{i} \right)a \left(t _{i} \right) \right]^{-1} 
= \frac{2}{3 \gamma  - 2}
\left[H \left(t _{e} \right)a \left(t _{e} \right) \right]^{-1}\ .
\label{22}
\end{equation}
Consequently, we obtain 
\begin{equation}
D _{\left(\nu \right)}^{^{\left(FLRW,d \right)}}\left(t _{e} \right) = 
\frac{1}{2}\frac{\left(3 \gamma- 2 \right)^{2}}
{3 \gamma + 2}
D _{\left(\nu \right)}^{^{\left(s d \right)}}\left(t _{i} \right) \ ,
\mbox{\ \ }
\left(\nu \ll 1 \right)\ . 
\label{23}
\end{equation}
For radiation ($\gamma  = 4/3$) the transfer function is  
$1/3$, while for dust ($\gamma = 1$) it is 
$1/10$. 

It is  
clear now, in which sense the exponentially decaying perturbations 
$D _{\left(\nu\right) }^{\left(s,d \right)}$ 
resurrect in order to yield the 
fluid energy density perturbations at reentry into the 
Hubble radius at $t = t _{e}$. 
Let us denote by $t _{f}$ the time at which the 
de Sitter phase finishes, i.e., 
$t _{i} < t _{f} < t _{e}$.  
During the time interval $t _{f} - t _{i}$ the scale factor increases 
exponentially with $H = {\rm const}$. 
Because of the $a ^{-2}$ dependence of the ``dominating'' mode, the fractional, comoving 
scalar field energy density perturbations are 
exponentially suppressed by 
many orders of magnitude, i.e.,   
$D _{\left(\nu\right) }^{\left(s,d \right)}
\left(t _{f} \right)$  is vanishingly small 
compared with 
$D _{\left(\nu\right) }^{\left(s,d \right)}\left(t _{i} \right)$. 
For $t > t _{f}$ the overall energy density perturbations are 
represented 
by the fluid quantity 
$D _{\left(\nu\right) }^{\left(FLRW,d\right)}$. 
Assuming an instantaneous transition to the radiation dominated 
FLRW stage, i.e., 
$D _{\left(\nu\right) }^{\left(s,d \right)}\left(t _{f} \right) 
\approx 
D _{\left(\nu\right) }^{\left(FLRW,d\right)}\left(t _{f} \right)$, 
we have $a \propto t ^{1/2}$ in a radiation dominated universe for $t > t _{f}$ and a 
power law 
growth $\propto a ^{2} \propto t$ of the quantity 
$D _{\left(\nu\right) }^{\left(FLRW,d\right)}$. 
Consequently, it needs a large time interval $t _{e} - t _{f}$ 
during which the power law growth of 
$D _{\left(\nu\right) }^{\left(FLRW,d \right)}$ compensates 
the initial exponential decay of 
$D _{\left(\nu\right) }^{\left(s,d \right)}$ during the interval 
$t _{f} - t _{i}$. 
Finally, at $t = t _{e}$, i.e. at the time at which the 
perturbation crosses the horizon inwards, the 
fluid energy density perturbations 
$D _{\left(\nu\right) }^{\left(FLRW,d\right)}$  are 
almost of the same order again as the scalar field perturbations 
$D _{\left(\nu\right) }^{\left(s,d\right)}$ at the first horizon 
crossing time $t _{i}$. 

Obviously, the description in terms of comoving energy density 
perturbations that initially become exponentially small but 
afterwards start to grow again is completely equivalent to the 
characterization with the help of conserved quantities.  
It is the advantage of our covariant approach that it is capable of 
providing us with a clear and transparent picture of the different 
aspects of the behaviour of large-scale perturbations in an 
inflationary universe. 
\section*{Acknowledgments}
This paper was supported by the Deutsche Forschungsgemeinschaft and  
NATO (grant CRG940598).

\end{document}